\documentclass[aps,prl,twocolumn,showpacs,preprintnumbers,floatfix,final]{revtex4}

\usepackage{graphicx,amsmath,amssymb,mathrsfs,array,longtable,delarray,verbatim,epstopdf}
\begin{document}

\title{Spin-incoherent transport in quantum wires}

\author{W. K. Hew}
\author{K. J. Thomas}
\author{M. Pepper}
\author{I. Farrer}
\author{D. Anderson}
\author{G. A. C. Jones}
\author{D. A. Ritchie}

\affiliation{Cavendish Laboratory, J. J. Thomson Avenue, Cambridge CB3 0HE, United Kingdom}


\begin{abstract}
When a quantum wire is weakly confined, a conductance plateau appears at $e^2/h$ in zero magnetic field accompanied by a gradual suppression of the $2e^2/h$ plateau with decreasing carrier density. Applying an in-plane magnetic field $B_{\parallel}$ does not alter the value of this quantisation; however, the $e^2/h$ plateau weakens with increasing $B_{\parallel}$ up to 9~T, and then strengthens on further increasing $B_{\parallel}$, which restores the $2e^2/h$ plateau. Our results are consistent with spin-incoherent transport in a one-dimensional wire.
\end{abstract}

 \maketitle

One of the major developments in the study of one-dimensional (1D) transport in the last decade was the discovery of the 0.7 structure\cite{thomas96a}. The 0.7 structure cannot be understood in the framework of one-electron theory and appears to be a consequence of electron spin. Interactions are believed to play a major role in 1D transport: a generic ground state is the Luttinger liquid, where the conductance is suppressed for repulsive interactions. However, it is now understood that charge interactions within the wire are masked in transport measurements, as the device is coupled to non-interacting Fermi-liquid leads where interactions cancel\cite{maslov95a,pmarenko95,safi95}. A conductance plateau at $2e^2/h$ is therefore observed at low temperatures irrespective of interaction strength\cite{tarucha95}. For low carrier densities, a 1D Wigner crystal with antiferromagnetic coupling is expected\cite{schulz93,glazman92,matveev04a}. Transport in this strongly-interacting r\'{e}gime remains largely unexplored, save in a few specific 2D systems\cite{glasson01, goldman90}.

One important consequence of electron-electron interaction is spin-charge separation, whereby spin and charge modes of the electron propagate at different velocities. Recent theoretical work\cite{fiete04,cheianov04,matveev04a,fiete07,matveev07a} explores the r\'{e}gime of the spin-incoherent Luttinger liquid, wherein the exchange coupling $J$ of neighbouring electrons becomes weak at low densities such that $J \ll k_{\mathrm{B}}T \ll E_{\mathrm{F}}$ and characteristics distinct from a normal Luttinger liquid are expected. Significantly, the spin modes are mostly reflected in such a r\'{e}gime, giving rise to an additional resistance contribution: conductance is suppressed by a factor of 2 to $e^2/h$\cite{matveev04b}. With strong exchange interaction, $J \gg k_{\mathrm{B}}T$, spin modes transport without scattering, the resulting charge modes giving rise to a conductance of $2e^2/h$, the usual Luttinger-liquid result.

In this Letter we show that, by tuning the confinement potential, a suppression of the $2e^2/h$ conductance plateau accompanied by a plateau at $e^2/h$ is observed as predicted in the spin-incoherent r\'{e}gime. In addition to the usual transport results of non-interacting 1D electrons, we report a transition into a strongly-interacting 1D state as the confinement potential is weakened\cite{meyer07}. A zero-bias anomaly (ZBA) peak, characteristic of ballistic 1D wires, is observed for conductances $G<e^2/h$; the peak does not split in $B_{\parallel}$, but is instead suppressed. We show that increasing $B_{\parallel}$ induces a spin incoherent-to-coherent transition in our system.

Split gates (width $0.7$~$\mu$m, length $0.4$~$\mu$m) were defined on a GaAs/AlGaAs single heterostructure 300~nm above the two-dimensional electron gas (2DEG), which had a mobility $1.85 \times 10^6$~cm$^2$/Vs and density $1.53 \times 10^{11}/$cm$^{2}$ after partial illumination. There was a top gate (width 1~$\mathrm{\mu}$m) above the split gates, separated by a layer of cross-linked polymethylmethacrylate (PMMA) 200~nm thick.\cite{{hew08}} Magnetic fields were applied with the sample aligned in-plane, parallel to the current direction. The out-of-plane misalignment was $1.2^{\circ}$, but the additional confinement in the $z$-plane does not affect the 1D dispersion appreciably.

Although the definition of a narrow channel in a 2DEG by means of split gates is a simple technique widely used to study ballistic transport in one-dimension\cite{thornton86}, the confinement potential is difficult to model. A voltage applied to the gates changes both the width and the carrier density of the 1D wire. Exploring conductances close to pinch-off, we can assume a parabolic confinement which become shallower as the width increases.\cite{laux88} With our geometry, we define the width of the wire by tuning the split-gate voltage $V_{\mathrm{sg}}$, and the top-gate voltage $V_{\mathrm{tg}}$ is then used to deplete the carriers. The two-terminal differential conductance $G=dI/dV_{\mathrm{sd}}$ was measured as a function of $V_{\mathrm{tg}}$ using a 33~Hz excitation voltage of 5 $\mu$V at a base electron temperature of some 100~mK.

\begin{figure}
\includegraphics[width=1.0\columnwidth]{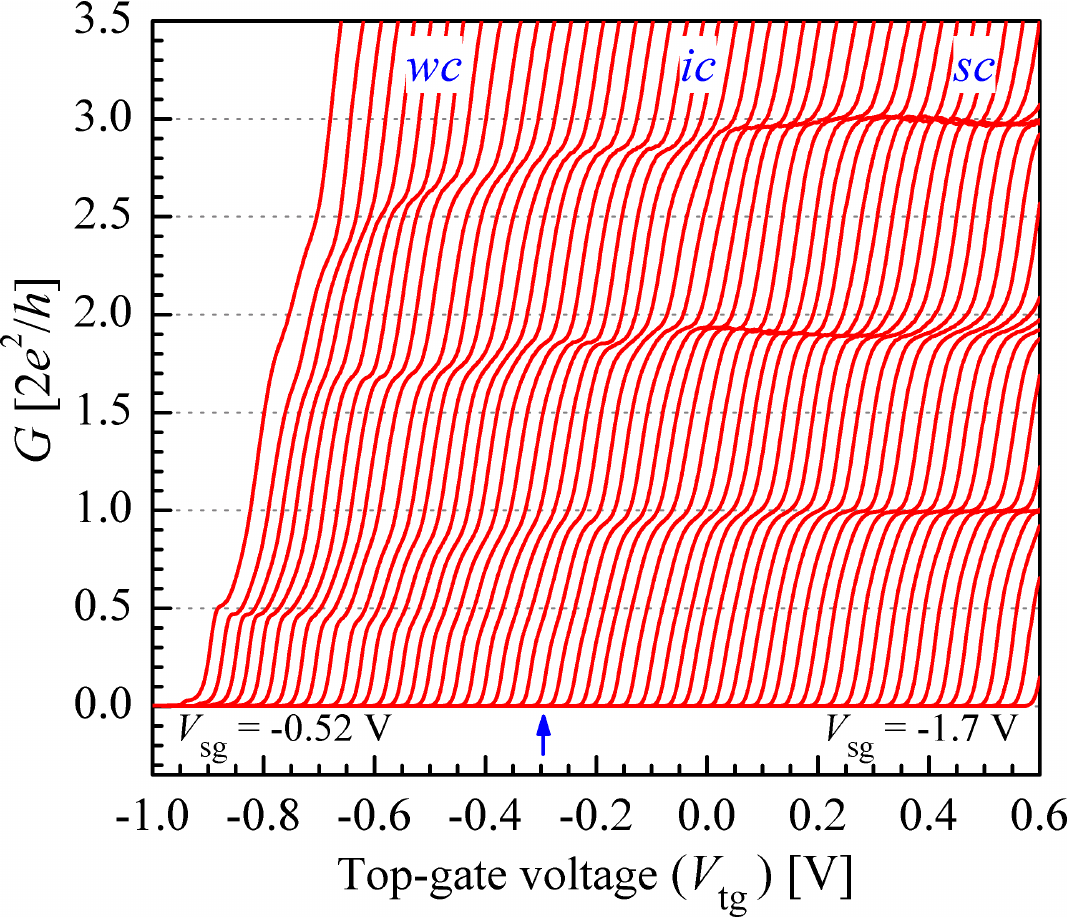}
\caption{Differential conductance $G(V_{\mathrm{tg}})$ measured at $T=50$~mK with fixed $V_{\mathrm{sg}}$, ranging from -1.7~V on the right-hand side to -0.52~V on the left hand side. A structure evolves at $e^2/h$ on the left-hand side accompanied by the suppression of the $2e^2/h$ degenerate plateau. Moving from right to left \textit{sc}, \textit{ic} and \textit{wc} mark the strong-, intermediate-, and weak-confinement r\'{e}gimes described in the text. The arrow at the bottom of the figure indicates the onset of a plateau at $e^2/h$ corresponding to the arrow shown in Fig. \ref{fig2}. \label{fig1}}
\end{figure}

It is difficult to measure precisely the electon density of a quantum wire, which hinders the quantitative linking of experimental results to theory, but we have nevertheless attempted to do so by transverse-field magnetic depopulation\cite{berggren86}, arriving at an estimated $n_{\mathrm{1D}} \sim 1 \times 10^{5}/$cm for single subband occupancy in the $wc$ r\'{e}gime (defined in Fig.~\,\ref{fig1}), satisfying $n_{\mathrm{1D}} \ll 1/a_{\mathrm{B}}$, where $a_B$ is the effective Bohr radius. With this density, we obtain $J/k_{\mathrm{B}} \sim 2.2$~mK and $E_{\mathrm{F}}/k_{\mathrm{B}} \sim 1.6$~K according to \cite{matveev04b}, which is consistent with the inequality $J \ll k_{\mathrm{B}}T \ll E_{\mathrm{F}}$.

\begin{figure}
\includegraphics[width=1.0\columnwidth]{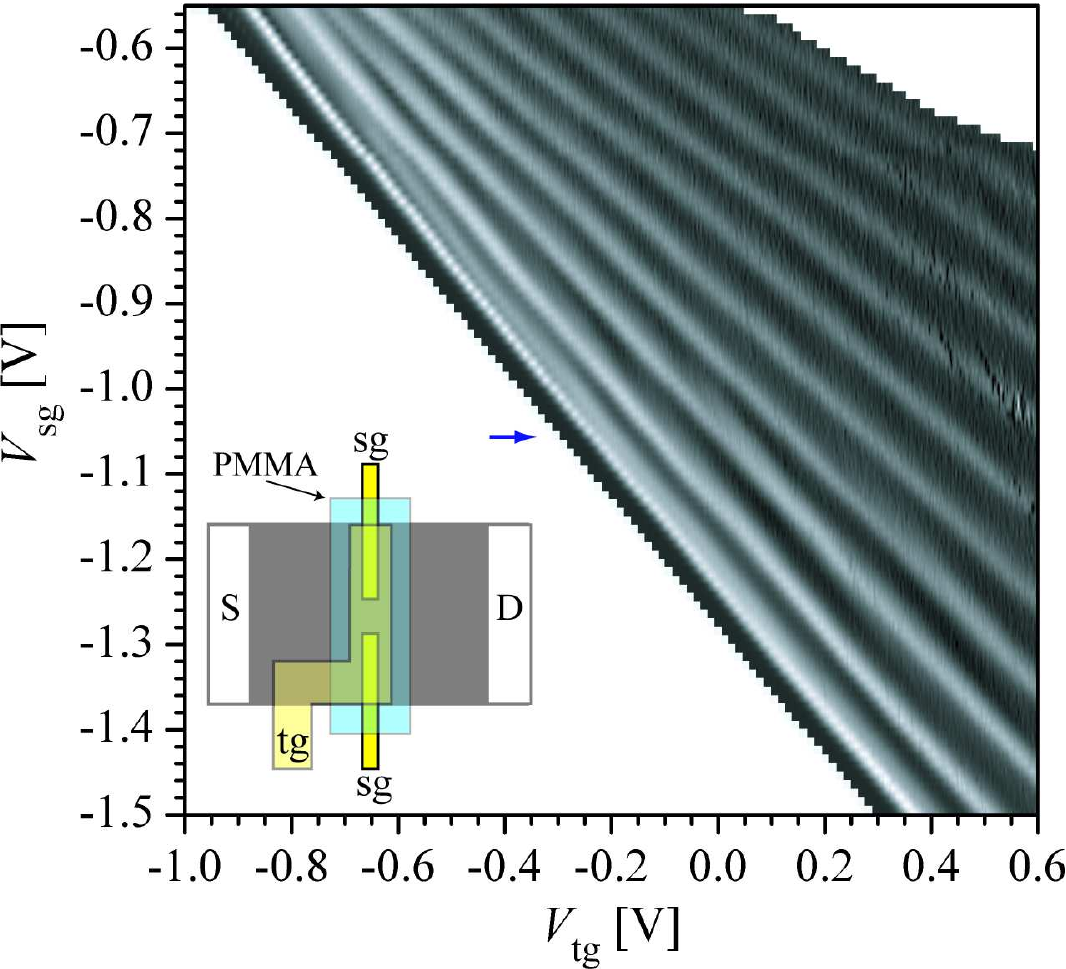}
\caption{Grey-scale plot of transconductance $dG/dV_{\mathrm{tg}}$. Gradual suppression of the $2e^2/h$ plateau and strengthening of the $e^2/h$ plateau are marked by a thin light line bridging the two adjacent light lines at the bottom. \textit{Inset}: Schematic diagram of the device (not to scale) showing the split gates (sg) and top gate (tg) separated by a dielectric layer (PMMA).
\label{fig2}}
\end{figure}

Figure\,\ref{fig1} shows differential conductance $G(V_{\mathrm{tg}})$ measured by sweeping $V_{\mathrm{tg}}$ at fixed  $V_{\mathrm{sg}}$. In general, moving from right to left, i.e., making $V_{\mathrm{sg}}$ more positive and $V_{\mathrm{tg}}$ more negative, the channel widens as confinement weakens, and carrier density decreases. The subband energies were measured (not shown) in different r\'{e}gimes: the spacing between the first and second subband decreases from $1.2$~meV to $0.7$~meV across the plot right-to-left. On the right, where $V_{\mathrm{sg}}$ is more negative, the plateaux are better quantized. This region is referred to as the strongly confined (\textit{sc}) r\'{e}gime in this letter. Moving to the left, we cross over to a weakly confined (\textit{wc}) r\'{e}gime through an intermediate confinement (\textit{ic}) r\'{e}gime. In the \textit{ic} r\'{e}gime, we already observe a gradual degradation and lowering of the $2e^2/h$ quantized plateau. The $2e^2/h$ quantized plateau disappears eventually in the \textit{wc} r\'{e}gime, leaving a strong structure at $e^2/h$. We note that this behaviour is only observed at low densities. Although the $2e^2/h$ plateau vanishes, suppressed higher-index plateaux persist in the \textit{wc} r\'{e}gime. This figure shows the three r\'{e}gimes with distinct characteristics. For a higher starting 2D carrier density, the \textit{wc} and \textit{ic} will show different behaviour. The suppression of the $2e^2/h$ plateau may still be observed, but is not accompanied by a $e^2/h$ plateau. As the density is increased, this distinction disappears, and \textit{sc} characteristics prevail throughout.

Figure\,\ref{fig2} shows a grey-scale plot of $dG/dV_{\mathrm{tg}}$ obtained by numerically differentiating $G(V_{\mathrm{tg}})$ for different $V_{\mathrm{sg}}$. In general, light lines correspond to the risers between plateaux and dark lines to the plateaux. Moving diagonally up, the dark lines shrink as $V_{\mathrm{sg}}$ increases, reflecting a shortening of plateaux or decreasing subband spacings. For higher-index subbands, the light and dark lines remain distinct for the full range of $V_{\mathrm{sg}}$. For the last subband, in addition to the shrinking dark line corresponding to the $2e^2/h$ plateau, a thin dark line appears at $V_{\mathrm{sg}}=-1.05$~V, marking the formation of a $e^2/h$ plateau. As $V_{\mathrm{sg}}$ is increased further, the thin line grows, eventually becoming the fully-developed $e^2/h$ plateau.

\begin{figure}
\includegraphics[width=1.0\columnwidth]{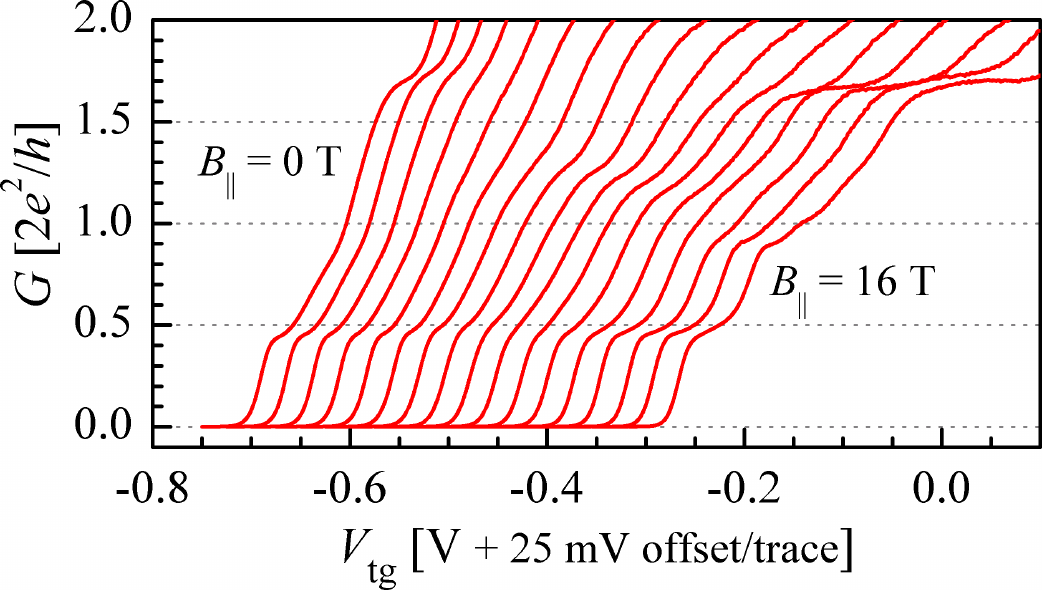}
\caption{Differential conductance $G(V_{\mathrm{tg}})$ in the \textit{wc} r\'{e}gime at fixed $B_{\parallel}$, incremented from 0 to 16 T in steps of 1~T. The $e^2/h$ plateau initially weakens up to 9~T, but no appreciable change in the quantized value is observed.
\label{fig3}}
\end{figure}

\begin{figure}
\includegraphics[width=1.0\columnwidth]{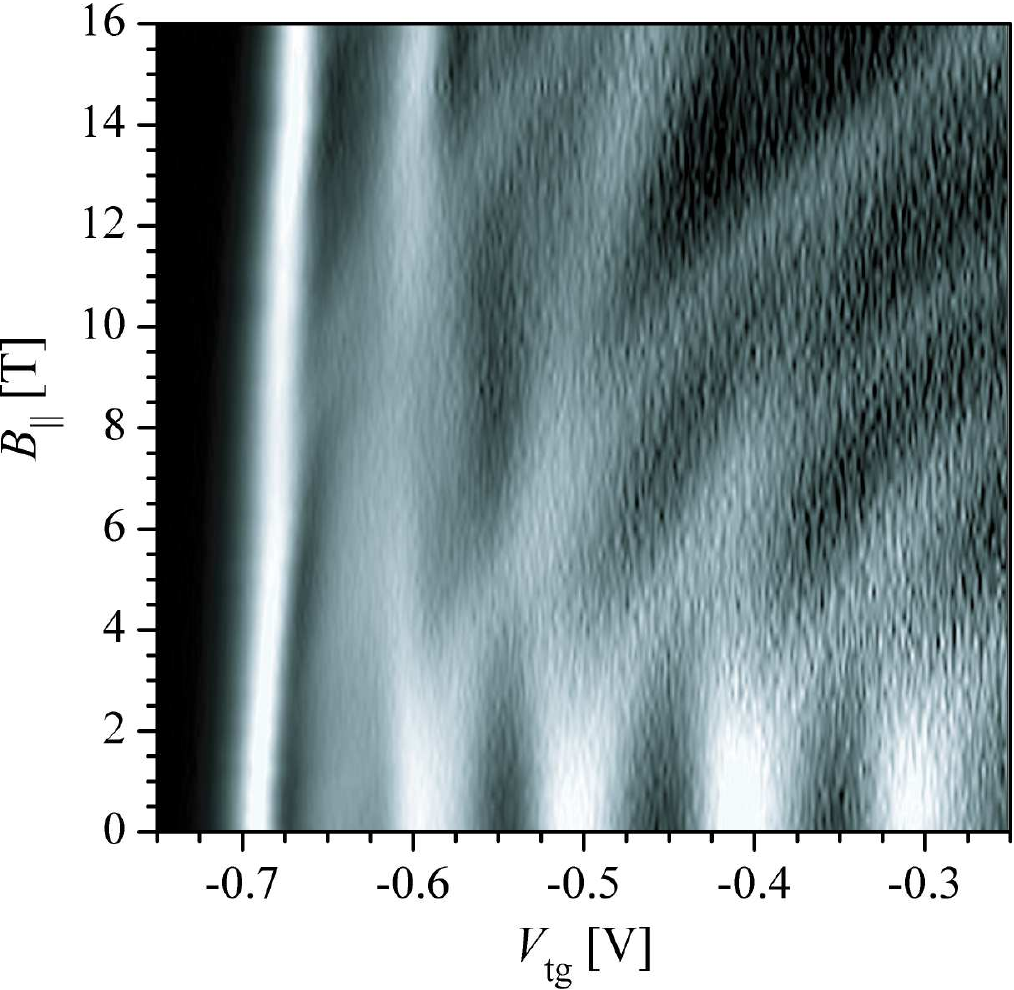}
\caption{Grey-scale plot of $dG/dV_{\mathrm{tg}}$ with $B_{\parallel}$. Diverging pairs of light lines for $V_{\mathrm{tg}}>-0.62$~V indicate Zeeman splitting for subbands $N \geq 2$. The sharp light line on the left-hand side at $V_{\mathrm{tg}}=-0.69$~V corresponds to spin-incoherent electrons. It appears to be the spin-split branch of $N=1$; however, note that at 9~T, a branch comes out to the right of this line, at which point the transport becomes spin-coherent.
\label{fig4}}
\end{figure}

Figure\,\ref{fig3} shows the magnetic-field dependence of the $e^2/h$ plateau. The plateau remains at $e^2/h$ with increasing $B_{\parallel}$, typically a sign of the ferromagnetic state. However, we see a weakening of the plateau as $B_{\parallel}$ increases to 9~T, which would not be expected of such a state. The plateau regains strength upon further increasing $B_{\parallel}$. The initial weakening of the plateau may be attributed to the decreasing stability of spin-incoherence with increasing field up to 9~T, whereupon the system undergoes a transition to spin coherence, and the strengthening plateau seen at higher fields is the usual spin-polarised plateau at $e^2/h$. It is noteworthy that the transition field of $B_{\parallel} \approx 9$~T corresponds to a Zeeman energy~$\sim 250$~$\mathrm{\mu}$eV. That such a large field, on the order of $E_{\mathrm{F}}$, is required to polarise the spins is quite remarkable.

A grey-scale plot of this data is shown in Fig.\,\ref{fig4}, where the Zeeman splitting of 1D subbands $N\geq2$ is characterized by diverging pairs of light lines from an apex at zero field. At $V_{\mathrm{tg}}=-0.69$~V, there is a sharp light line and at $V_{\mathrm{tg}}=-0.65$~V, there is a broad ``foggy" line: the former corresponds to the riser to the $e^2/h$ plateau and the latter to a shallow change of slope between $e^2/h$ and $2e^2/h$. The sharp light line bifurcates as $B_{\parallel}$ increases beyond 9~T, representing spin coherence with Zeeman splitting. In strongly-confined 1D wires, Zeeman splitting results in diverging pairs of transconductance peaks with a \textbf{V} shape rather than the \textbf{Y} shape we see here. The stem of the \textbf{Y} shrinks as $N$ increases, thereby reducing the threshold for transition into coherent transport. For large $B_{\parallel}$, spin is well defined and the $2e^2/h$ plateau appears at 14~T, whereas farther down the stem, spin becomes increasingly ill defined and eventually the incoherent r\'{e}gime takes over, with a reduction in conductance to $e^2/h$. It is difficult to infer what happens to the broad foggy line that lies close to the spin down (${\downarrow}$, low energy) branch of $N=2$ subband. It appears to merge with the $N=2$ spin-down branch, evolving together thereafter.

In previous measurements\cite{thomas00,pyshkin,reilly02,crook06}, a zero-field structure at $e^2/h$ was ascribed to spin polarisation. We note that, in these results, a structure at $e^2/h$ was also accompanied by a plateau at $2e^2/h$, which means a diverging pair of transconductance peaks corresponding to spin $\downarrow$ and spin $\uparrow$ with $B_{\parallel}$. We measured several samples in the current study, and, in all cases, the structure at $e^2/h$ was observed at low densities and weak confinement strengths.

Although it has been predicted that the ballistic quantisation of conductance in a quantum wire may be lost at low densities if the Fermi wavelength becomes comparable with the wire length\cite{glazman88}, the restoration of the quantised plateau at $2e^2/h$ in a high in-plane magnetic field reassures us that this is not the case here. The constancy of the plateau at $e^2/h$ indicates that transmission is near unity as a high $B_{\parallel}$ would otherwise halve its value.

The simultaneous weakening of the $2e^2/h$ plateau with the appearance of an $e^2/h$ plateau and the unusual magnetic field dependence of the $e^2/h$ plateau give us confidence that our results differ from that of a spin-polarized state. We believe that this is a clear signature of suppression of the spin modes, and our system has attained the spin-incoherent Luttinger-liquid r\'{e}gime. We also stress that there is no degradation of the higher-index plateaux.

\begin{figure}
\includegraphics[width=1.0\columnwidth]{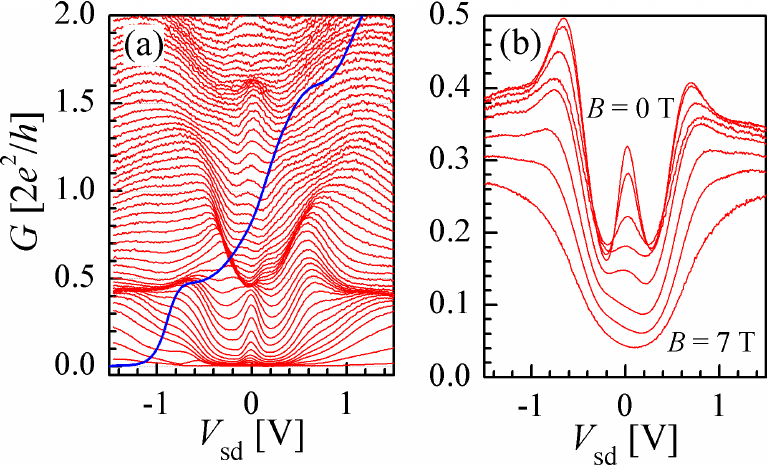}
\caption{\textbf{(a)} $G(V_{\mathrm{sd}})$ showing a Zero-Bias Anomaly peak for $G<e^2/h$ and its suppression above $e^2/h$. \textbf{(b)} $G(V_{\mathrm{sd}})$ measured at fixed $B_{\parallel}$. The ZBA shows no discernible splitting up to $B_{\parallel}=7$~T, beyond which the peak is completely suppressed.
\label{fig5}}
\end{figure}

\begin{figure}
\includegraphics[width=1.0\columnwidth]{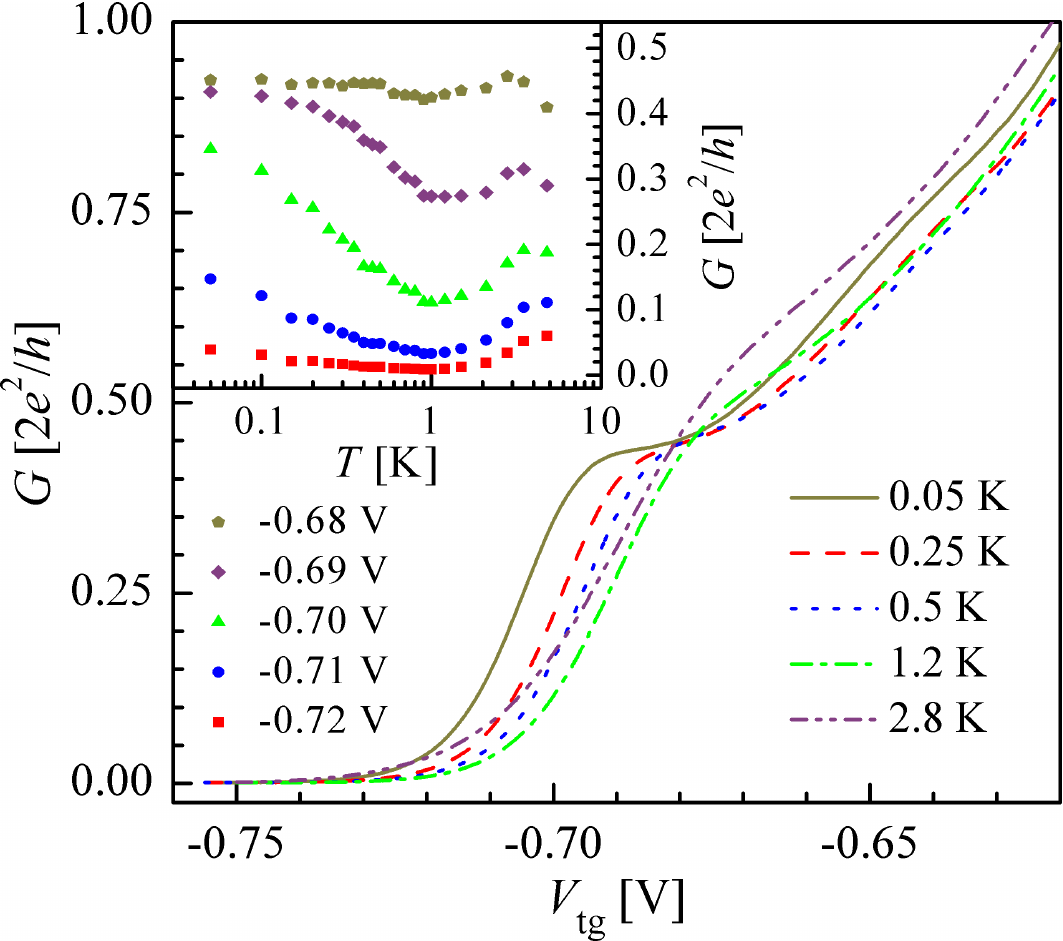}
\caption{Dependence of the $e^2/h$ plateau on the lattice temperature. There is overall thermal averaging, a 0.7 structure evolves at higher $T$. \textit{Inset}: Temperature dependence of $G$ at various carrier densities ($V_{\mathrm{tg}}$); the top trace is for the midpoint of the plateau.
\label{fig6}}
\end{figure}

Figure \ref{fig5}(a) shows $G(V_{\mathrm{sd}})$ with well-defined ZBA conductance peaks for $G<e^2/h$ and significantly different behaviour for $G>e^2/h$. The magnetic-field dependence of the ZBA is shown in figure \,\ref{fig5}(b); unlike previous other work\cite{cronenwett02}, there is no splitting of the ZBA with increasing $B_{\parallel}$, the peak fully suppressed above 7~T. Figure\,\ref{fig6} shows the temperature dependence of the $e^2/h$ plateau, which thermally smears as $k_{\mathrm{B}}T$ approaches the Fermi energy, whilst a 0.7 structure appears at $T>1$~K. A previous study of a spin-polarized $e^2/h$ state in a magnetic field showed a continuous rise in the conductance structure at $e^2/h$ towards a 0.7 structure with increasing $T$ \cite{thomas02}. The present result is different in this respect as well: inset in Fig. \ref{fig6} are detailed plots of the temperature dependence at various fixed $V_{\mathrm{tg}}$; there is an initial decrease in conductance observed for $T<1$~K, above which it rises, in agreement with the evolution of 0.7 structure at higher $T$. It is noteworthy that the 0.7 structure emerges without an accompanying ZBA, nor is there a low-temperature plateau at $2^2/h$ from which it evolves. Moreover, the absence of Zeeman splitting in the ZBA peak, quite possibly a signature of spin-incoherent transport in our system, discounts the Kondo effect as a possible explanation.

As a final remark, we note that a small transverse field $B_z=0.35$~T re-establishes the $2e^2/h$ plateau. It is possible that the transverse field provides additional confinement, Lorentz force separating electrons of opposite momenta to the channel edges, restoring spin coherence.

In conclusion, a quantum wire at low carrier density shows a conductance plateau at $e^2/h$ in zero magnetic field, accompanied by a suppression of the $2e^2/h$ plateau. As the confinement is strengthened, the standard $2e^2/h$ quantization of conductance (as described by the non-interacting model) is recovered, the r\'{e}gime in which quantum wires have largely been measured by us and by others previously. On the other hand, the low-density effects reported here are consistent with the predictions of spin-incoherent Luttinger-liquid theory.

We thank K.~A. Matveev for stimulating discussions. This work was supported by the EPSRC. WKH acknowledges the Cambridge Commonwealth Trust and St.~John's College Benefactors' Scholarships, and KJT the Royal Society Research Fellowship.

\bibliographystyle{apsrev}

\end{document}